\documentclass{PoS}
\usepackage{txfonts}

\title{Exploring the faint source population at 15.7 GHz}

\ShortTitle{Exploring the faint source population at 15.7 GHz}

\author{\speaker{Imogen H. Whittam}\\
        Physics and Astronomy Department, University of the Western Cape, Bellville 7535, South Africa\\
        E-mail: \email{imogenwhittam@gmail.com}}

\author{Julia M. Riley\\
        Astrophysics Group, Cavendish Laboratory, 19 J.~J.~Thomson Avenue, Cambridge CB3 0HE, UK}

\author{David A. Green\\
        Astrophysics Group, Cavendish Laboratory, 19 J.~J.~Thomson Avenue, Cambridge CB3 0HE, UK}

\author{Matt J. Jarvis\\
        Astrophysics, University of Oxford, Denys Wilkinson Building, Keble Road, Oxford, OX1 3RH, UK\\
        Physics and Astronomy Department, University of the Western Cape, Bellville 7535, South Africa}

\abstract{We discuss our current understanding of the nature of the faint, high-frequency radio sky. The Tenth Cambridge (10C) survey at 15.7 GHz is the deepest high-frequency radio survey to date, covering 12 square degrees to a completeness limit of 0.5 mJy, making it the ideal starting point from which to study this population. In this work we have matched the 10C survey to several lower-frequency radio catalogues and a wide range of multi-wavelength data (near- and far-infrared, optical and X-ray). We find a significant increase in the proportion of flat-spectrum sources at flux densities below 1~mJy -- the median radio spectral index between 15.7~GHz and 610~MHz changes from 0.75 for flux densities greater than 1.5~mJy to 0.08 for flux densities less than 0.8~mJy. The multi-wavelength analysis shows that the vast majority (> 94 percent) of the 10C sources are radio galaxies; it is therefore likely that these faint, flat spectrum sources are a result of the cores of radio galaxies becoming dominant at high frequencies.
We have used new observations to extend this study to even fainter flux densities, calculating the 15.7-GHz radio source count down to 0.1 mJy, a factor of five deeper than previous studies. There is no evidence for a new population of sources, showing that the high-frequency sky continues to be dominated by radio galaxies down to at least 0.1~mJy.}

\FullConference{EXTRA-RADSUR2015 ($\star$)\\
		20--23 October 2015\\
		Bologna, Italy

                \bigskip
                \hrule
                \bigskip

                \textnormal{($\star$) This conference has been organized
                  with the support of the Ministry of Foreign Affairs
                  and International Cooperation, Directorate General
                  for the Country Promotion (Bilateral Grant Agreement
                  ZA14GR02 - Mapping the Universe on the Pathway to
                  SKA)}
}

\begin{document}

\section{Introduction}

The high frequency ($\gtrsim 10$~mJy) extragalactic radio sky has been much less widely studied than that at lower frequencies (e.g.\ 1.4~GHz), mostly due to the increased telescope time required to conduct a survey to an equivalent depth at higher frequencies. As a result, while the nature of the low-frequency extragalactic source population is well constrained down to the micro-Jansky level (e.g.\ \cite{2009ApJ...694..235P,2010A&ARv..18....1D,2013MNRAS.436.1084M}), the composition of the higher-frequency sky is relatively unknown. 

Two surveys have probed this population at higher flux densities; the Ninth Cambridge (9C, \cite{2003MNRAS.342..915W,2010MNRAS.404.1005W}) survey, complete to 25~mJy, and the Australia Telescope 20~GHz (AT20G; \cite{2010MNRAS.402.2403M}) survey, complete to 40~mJy. These surveys show that the high-frequency sky is dominated by radio-loud sources at $S \gtrsim 10$~mJy. 

There have been several attempts to model the high-frequency source population (e.g.\ \cite{2005A&A...431..893D,2008MNRAS.388.1335W}), often extrapolating from lower frequencies. These models, however, are increasingly poor fits to the observed source counts, significantly underestimating the number of sources below 10~mJy \cite{2011MNRAS.415.2708A}. This demonstrates that the high-frequency extragalactic source population is poorly understood, partly due to the complexity and diversity of the high-frequency spectra of individual sources.

There is therefore a clear need for a detailed, multi-frequency study of the mJy and sub-mJy high-frequency sky. The Tenth Cambridge (10C; \cite{2011MNRAS.415.2699A,2011MNRAS.415.2708A}) survey provides the ideal starting point for just such a study. The 10C survey was observed with the Arcminute Microkelvin Imager (AMI; \cite{2008MNRAS.391.1545Z}) telescope in Cambridge, UK and covers a total of 27~deg$^2$ complete to 1~mJy and a further 12~deg$^2$, contained within these fields, complete to 0.5~mJy. Recently, new observations within two of these fields have extended the 10C survey down to 0.1~mJy \cite{2016arXiv160100282W}.

In the first part of these proceedings (Sections~\ref{section:radio} and \ref{section:multi}) we discuss the results of a detailed study of a sample selected from the 10C survey in the Lockman Hole field, a region which has been observed at a wide range of frequencies. This work is described in detail in \cite{2013MNRAS.429.2080W} and \cite{2015MNRAS.453.4244W}.  We then go on to present (Section~\ref{section:10Ccont}) the results of the recent deeper extension to the 10C survey \cite{2016arXiv160100282W} which provides further insights into the nature of the faint, high-frequency sky. In Section~\ref{section:models} we compare these observations to several models of the extragalactic source population.

\section{Radio properties}\label{section:radio}

\begin{figure}
\centerline{\includegraphics[width=7cm]{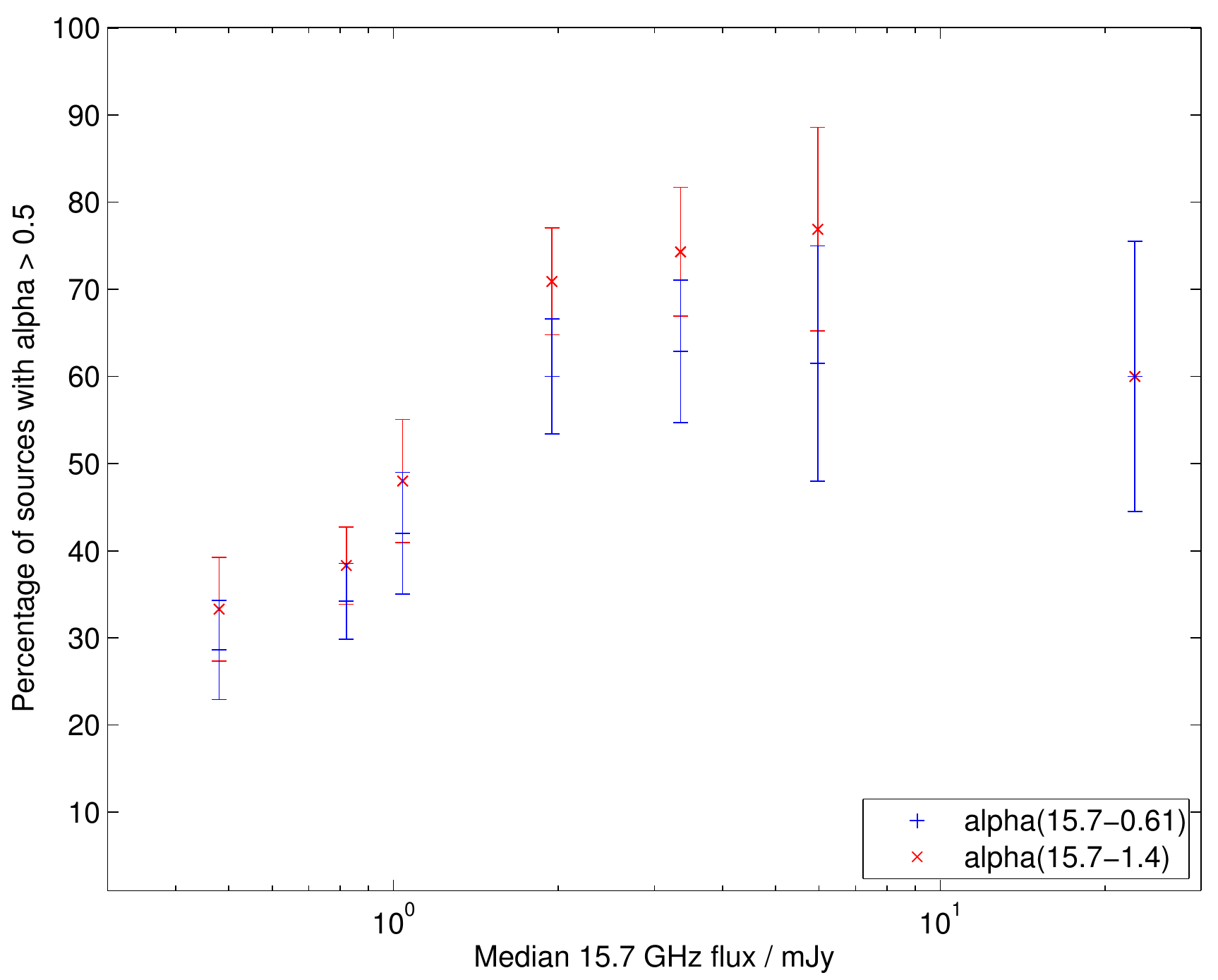}}
\caption{Percentage of steep spectrum sources (sources with $\alpha > 0.5$) in a range of 15.7~GHz flux density bins.}\label{fig:alpha}
\end{figure}

A sample of 296 sources was selected from the two fields of the 10C survey in the Lockman Hole. To investigate the radio properties of these 15.7-GHz-selected sources, the sample was matched to several lower-frequency radio catalogues available in the field. These are: a 610-MHz Giant Metrewave Radio Telescope (GMRT) catalogue \cite{2008MNRAS.387.1037G,2010BASI...38..103G}, a 1.4-GHz Westerbork Synthesis Radio Telesecope (WSRT) catalogue \cite{2012rsri.confE..22G}, two deep 1.4-GHz Very Large Array (VLA) catalogues at 1.4~GHz \cite{2006MNRAS.371..963B,2008AJ....136.1889O}, the Faint Images of the Radio Sky at Twenty-one centimetres (FIRST; \cite{1997ApJ...475..479W}) and the National Radio Astronomy Observatory (NRAO) VLA Sky Survey (NVSS; \cite{1998AJ....115.1693C}). Full details of the method used to match the catalogues can be found in \cite{2013MNRAS.429.2080W}. 

To help to understand the nature of the 10C sources radio spectral indices $\alpha$ (where $S \propto \nu^{-\alpha}$ for flux density $S$ and frequency $\nu$) are calculated for all 296 sources in the sample. 30 of the sources have no lower-frequency counterparts available, so an upper limit is placed on their spectral index. We find a significant change in the spectral index distribution of the sample with flux density; the proportion of flat spectrum sources increases significantly below 1~mJy, as shown by Fig.~\ref{fig:alpha}.  Further illustrating this point, the median radio spectral index between 15.7 GHz and 610 MHz changes from 0.75 for flux densities greater than 1.5 mJy to 0.08 for flux densities less than 0.8 mJy. 
There is clearly a population of faint, flat-spectrum sources emerging below 1~mJy.

\section{Multi-wavelength properties}\label{section:multi}

To help elucidate the nature of these flat-spectrum sources a range of multi-wavelength data are used. For this study, we use a complete sub-sample of 96 of the 10C sources studied in Section~\ref{section:radio}. This sample is matched to the Data Fusion catalogue compiled by Vaccari et al. (in preparation, \texttt{www.mattiavaccari.net/df}), which contains most of the publicly available photometry and spectroscopy for sources detected in the \emph{Spitzer} Extragalactic Representative Volume Survey (SERVS; \cite{2012PASP..124..714M}). As well as the 3.6~$\muup$m and 4.5~$\muup$m SERVS data, the Data Fusion catalogue includes optical photometry (GS11; \cite{2011MNRAS.416..927G}), near-infrared photometry from the United Kingdom Infrared Telescope (UKIRT) Infrared Deep Sky Survey  (UKIDSS; see \cite{2007MNRAS.379.1599L}) and mid-infrared and far-infrared photometry from the Spitzer Wide-Area Infrared Extragalactic survey (SWIRE; see \cite{2003PASP..115..897L}).  Multi-wavelength counterparts are found for 80 out of the 96 sources. Spectroscopic redshifts are available for 24 sources, and photometric redshifts are computed for a further 54 sources. Full details of the catalogues used, the methods used to identify the counterparts and calculate the photometric redshifts are described in \cite{2015MNRAS.453.4244W}.

A key aim of this work is to establish the nature of the flat spectrum sources which we have found to dominate the high-frequency population below $\sim 1$~mJy. Radio-to-optical and radio-to-infrared ratios ($R$ and $R_{3.6~\rm \muup m}$) are therefore a valuable tool, as they allow radio-loud AGN to be separated from radio-quiet AGN and star-forming galaxies. The two ratios\footnote{$R = S_{1.4~\rm GHz} \times 10^{0.4(m-12.5)}, \quad {\rm where}~m~{\rm is~the~optical~magnitude~in~the}~i{\rm -band}$, and $R_{3.6~\rm \muup m} = \frac{S_{1.4~\rm GHz}}{S_{3.6~\rm \muup m}}$} are calculated for the 96 10C sources in this sample. Sources with $R > 1000$ or $R_{3.6~\rm \muup m} > 3.1$ are considered radio loud, while sources with ratios less than these values are classified as radio quiet, and are therefore most likely either star-forming galaxies or radio-quiet AGN.

\begin{figure}
\centerline{\includegraphics[width=7cm]{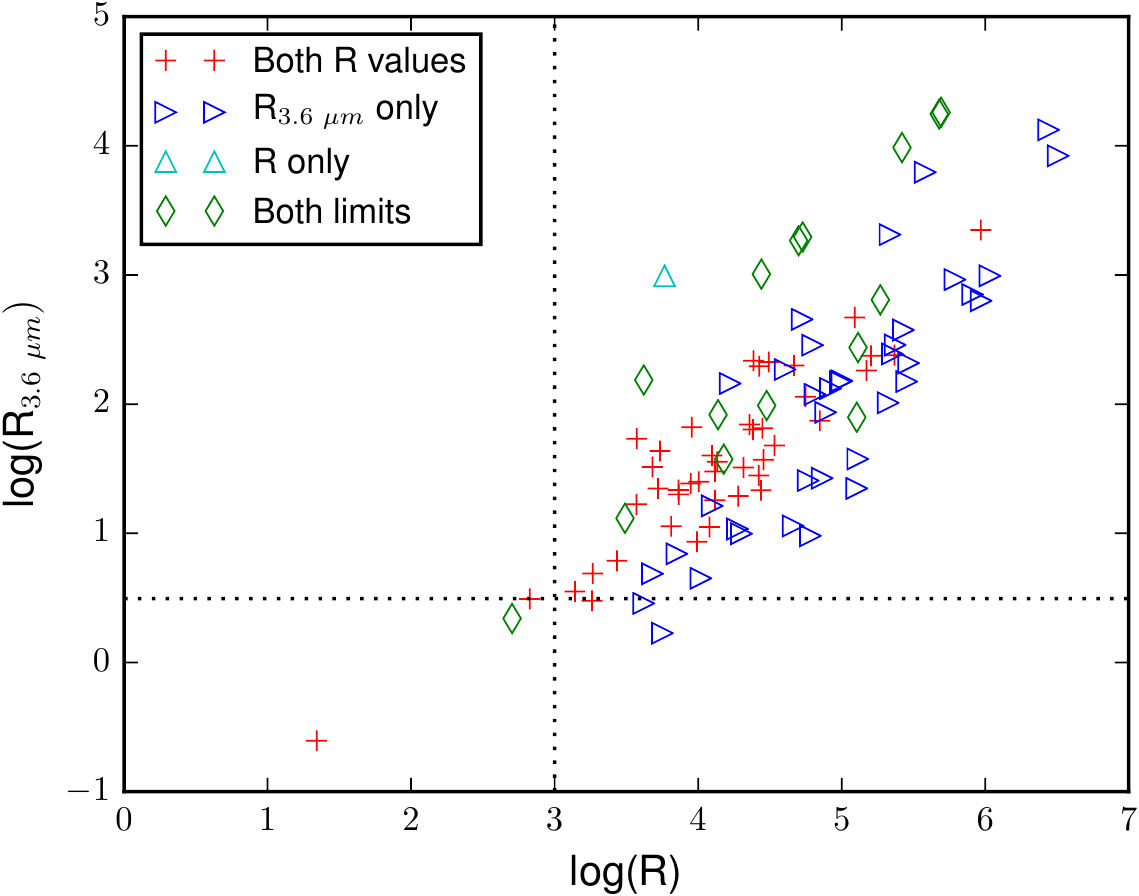}}
\caption{Radio-to-optical ($R$) and radio-to-infrared ($R_{3.6~\rm \muup m}$) values for 96 sources. Sources with only a radio-to-infrared $R_{3.6~\rm \muup m}$ value have a lower limit on $R$ and could therefore move to the right. Sources which only have an optical $R$ value could move up. Sources with a lower limit on both values of $R$ and $R_{3.6~\rm \muup m}$ could move up or to the right. The values of $R$ and $R_{3.6~\rm \muup m}$ used to distinguish between radio-quiet and radio-loud sources are shown by the dotted lines.}\label{fig:R}
\end{figure}

The radio-to-optical and radio-to-infrared ratios, along with the values used to classify the sources, are shown in Fig.~\ref{fig:R}. It is clear from this figure that the vast majority of the 10C sources are radio loud; only one source is clearly radio quiet. A further five sources could be classified as radio quiet using one or other of the $R$ values; however, given how close to the boundary they lie it is likely that they are also dominated by AGN activity. Therefore at least 90/96 (94 per cent) of the 10C sample are radio-loud sources. This shows that the the flat-spectrum sources discussed in Section~\ref{section:radio} are radio-loud AGN, most likely dominated by flat-spectrum self-absorbed synchrotron emission from their cores. Further work has shown that these sources are mostly compact radio galaxies, and are a mixed population of high-excitation and low-excitation (HERGs and LERGs) sources (Whittam et al., in prep.). In this work we compare the properties of these high-frequency HERGs and LERGs.

\section{10C extension}\label{section:10Ccont}

Recent observations over a total of 0.56~deg$^2$ in two of the 10C fields (the Lockman Hole field, J1052+5730, and the AMI001 field, J0024+3152) have allowed us to extend this study of the high-frequency extragalactic source population to even fainter flux densities. These data, which have a best rms noise of 16~$\mu$Jy beam$^{-1}$ and are fully described in \cite{2016arXiv160100282W}, enable us to probe the 15.7-GHz source population down to 0.1~mJy, a factor of five deeper than the existing 10C survey. 

\begin{figure}
\centerline{\includegraphics[width=7cm]{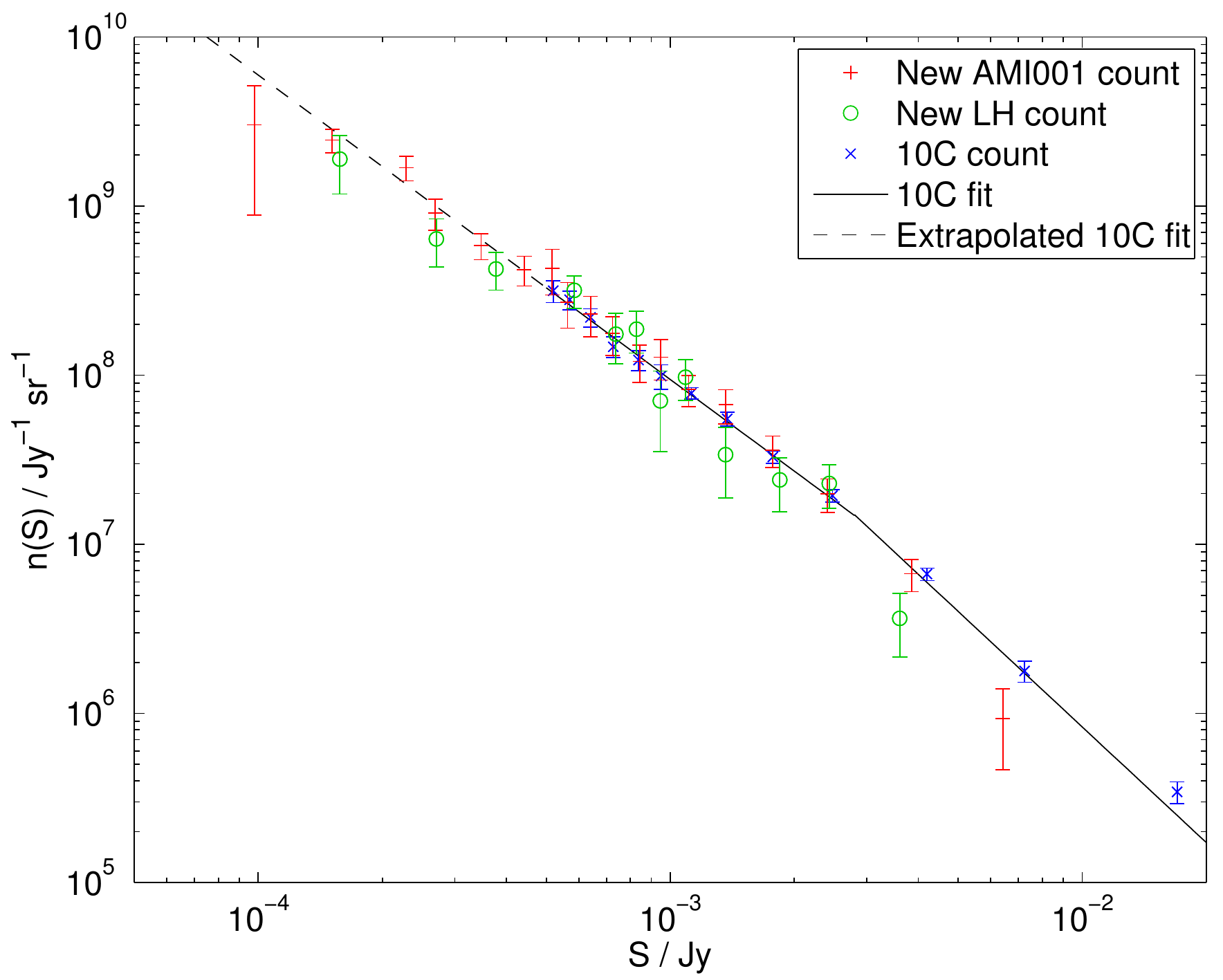}}
\caption{Source counts from the new observations and the original 10C survey. The new counts are in two fields: the AMI001 field (shown by red `+') and Lockman Hole field (green `$\circ$'). The original 10C source counts are shown as blue `$\times$'. The black lines are the broken power-law fitted to the original 10C counts. The faintest bin plotted for the AMI001 field count is based on only three sources and the completeness correction is not well defined at this flux density level, so this point is not included in the discussion or subsequent plots.}\label{fig:counts}
\end{figure}

The 15.7-GHz source count is shown in Fig.~\ref{fig:counts}. The new deeper source counts are consistent with the extrapolated fit to the 10C source count and display no evidence for a steepening or flattening of the count. There is therefore no evidence for a new population of objects contributing to the high-frequency sky at flux density levels above 0.1~mJy, suggesting that the source population continues to be dominated by radio galaxies. 

\section{Comparison with models}\label{section:models}

\begin{figure}
\centerline{\includegraphics[width=7cm]{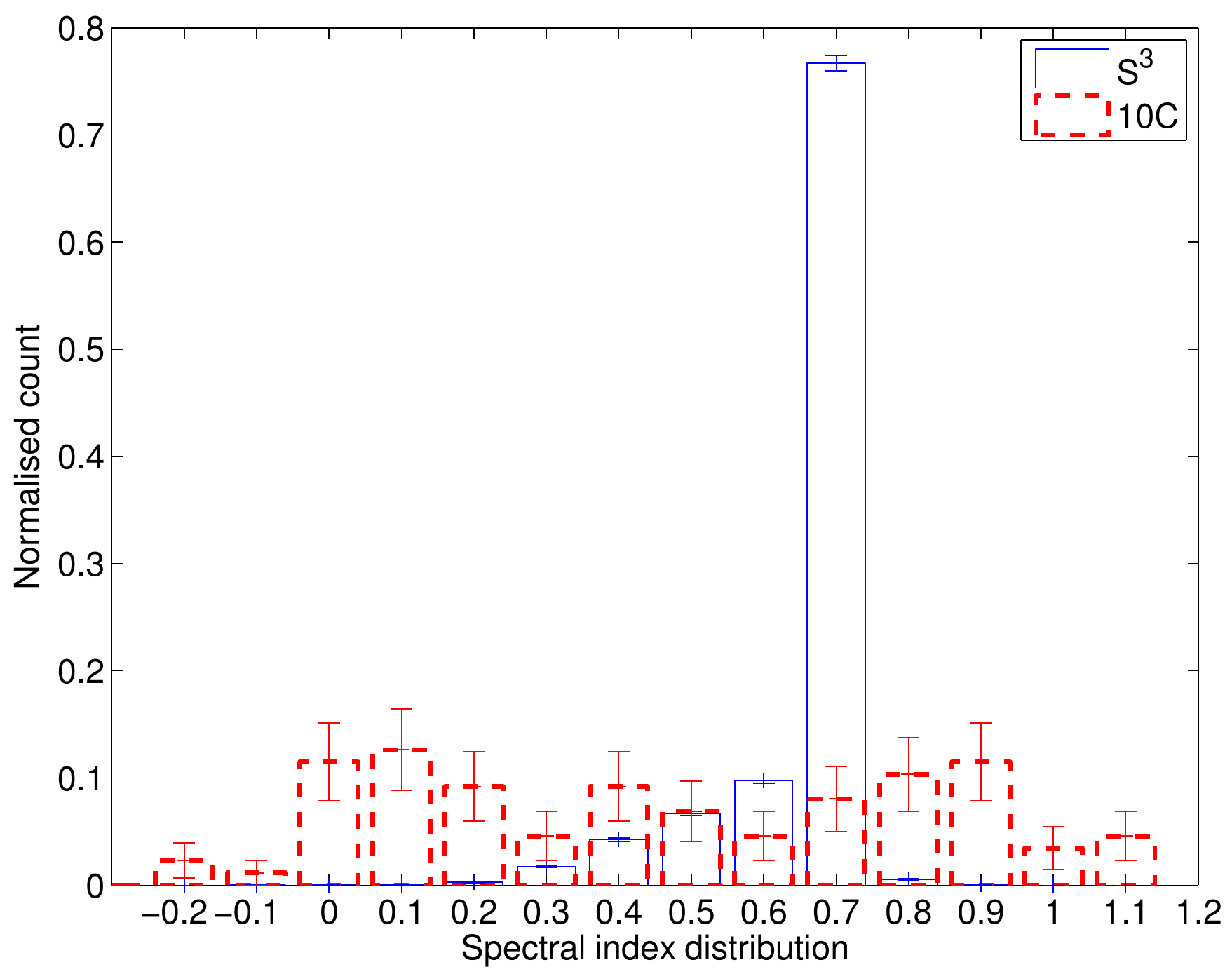}
\quad
\includegraphics[width=7cm]{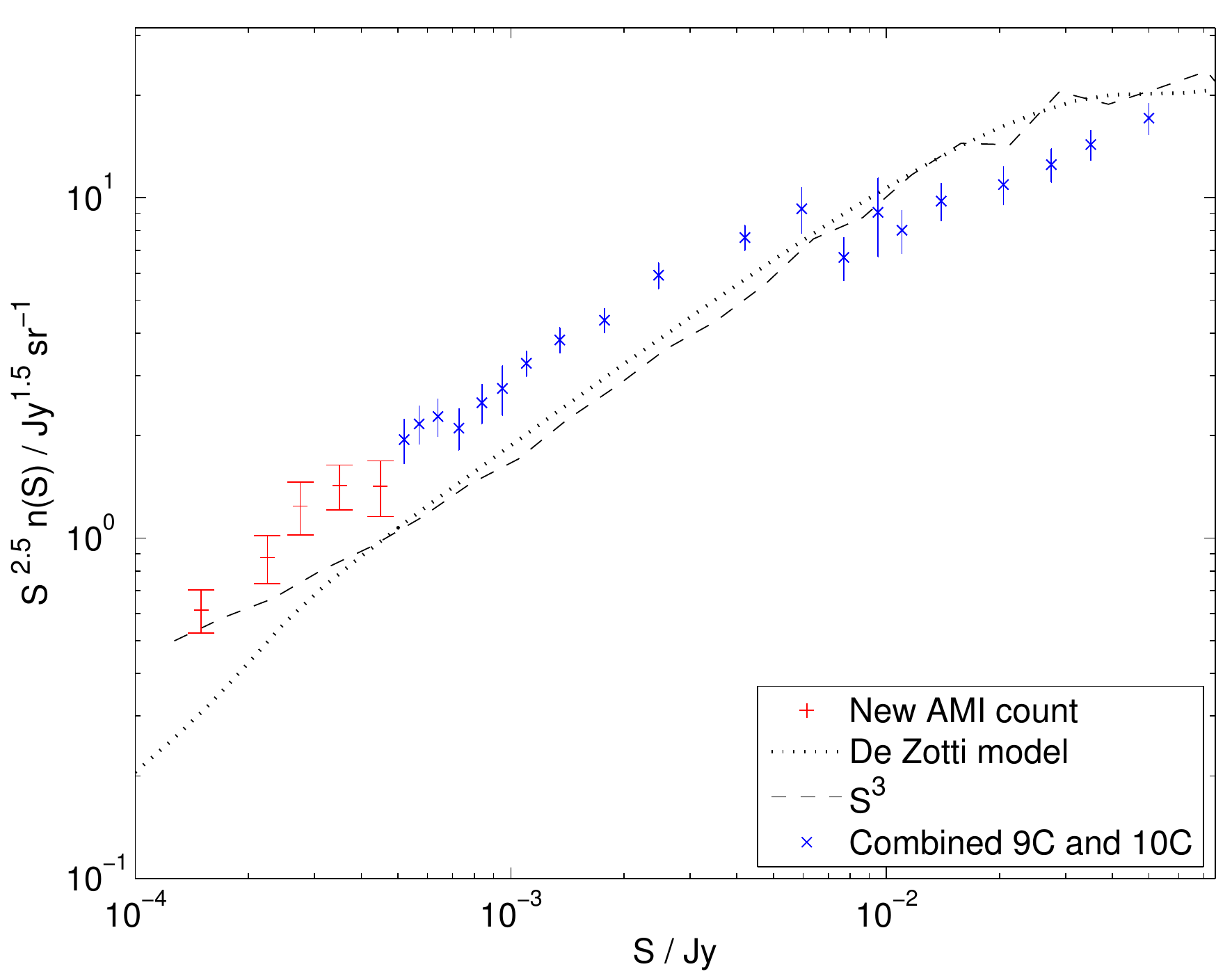}}
\caption{Comparisons between the observed 10C sample and the SKADS Simulated Skies. The left hand panel shows the spectral index distribution from the full Lockman Hole 10C sample and that predicted by the SKADS Simulated Skies. The right hand panel shows the Euclidean normalised differential source counts from the new 10C observations (from both fields combined, red `+') and the combined 9C and 10C source counts (blue `$\times$'). The de Zotti model \cite{2005A&A...431..893D} at 15~GHz (dotted line) and the 18~GHz count from the S$^3$ catalogue (dashed line) are also shown (no attempt is made to correct the latter to 15~GHz). Poisson errors are plotted for the observed counts.}\label{fig:s3-alpha}
\end{figure}

The Square Kilometre Array Design Study (SKADS) Simulated Skies (S$^3$; \cite{2008MNRAS.388.1335W,2010MNRAS.405..447W}) is a semi-empirical simulation of the extragalactic radio sky which covers a sky area of $20 \times 20~\rm deg^2$ out to a cosmological redshift of $z = 20$, and down to flux density limits of 10~nJy at 151, 610 MHz 1.4, 4.86 and 18 GHz. 
The simulation predicts that the 10C sample is dominated by Fanaroff--Riley type I (FRI, \cite{1974MNRAS.167P..31F}) sources (71 per cent), followed by FRII sources (13 per cent), with a smaller contribution from radio-quiet AGN and star-forming galaxies (together making up 10 per cent). 
The spectral index distribution predicted by the simulation is compared with the observed distribution from the 10C sample in the left hand panel of Fig.~\ref{fig:s3-alpha}. It is immediately clear that the two distributions are significantly different; the simulated distribution shows a narrow peak at $\alpha \approx 0.7$, while the 10C sources show a much flatter distribution, covering the full range of spectral indices $-0.2 < \alpha < 1.1$. The failure of the simulation to reproduce the observed distribution at $\alpha > 0.8$ is unsurprising, as there is an input assumption in the simulation that all the extended emission has $\alpha=0.75$ and no attempts are made to model spectral ageing. More surprising is the lack of flat-spectrum sources in the simulation; there are essentially no simulated sources with $\alpha < 0.3$, while a significant number of the observed sources do have spectra flatter than 0.3. This is probably because the emission from the cores of the FRI sources is much more dominant than the model suggests, causing these sources to have much flatter spectra than predicted.

The source count from the 10C survey is compared to the S$^3$ source count and that predicted by the de Zotti et al.\ \cite{2005A&A...431..893D} model in the right hand panel of Fig.~\ref{fig:s3-alpha}. Both models, which are derived empirically using lower-frequency data, underestimate the number of sources between 0.1 and 5~mJy by a factor of two. The number of flat spectrum sources in particular is underestimated; the de Zotti et al.\ model predicts that at $S_{15~\rm GHz} = 1~\rm mJy$ steep-spectrum sources outnumber flat-spectrum source by a factor of three, while the observations show that there are twice as many flat-spectrum sources as steep-spectrum sources. This is likely to be because both models fail to include the dominance of the flat-spectrum cores of radio galaxies and the relative faintness of the extended emission at high frequencies. 

The S$^3$ count shows a flattening at $\approx 0.3$~mJy, due to an increased contribution from star-forming galaxies (which comprise 20 per cent of the simulated population in the range $0.1 < S/\rm mJy < 0.3$). However, this flattening is not seen in the deep 10C count so it is not clear what contribution, if any, star-forming galaxies make to the source population at $S_{15.7} > 0.1~\rm mJy$. This is consistent with several recent studies of the faint ($S_{1.4~\rm GHz} < 0.1$~mJy) source population at lower frequencies \cite{2012MNRAS.421.3060S,2014MNRAS.440.1527L,2015AJ....150...87L}, which also find fewer star-forming galaxies than predicted by the simulation.

\section{Conclusions}\label{section:conclusions}

We have used a range of radio and multi-wavelength data to study a sample of sources selected at 15.7~GHz in the Lockman Hole. We find a population of flat-spectrum sources emerging below 1~mJy which is not predicted by models of the high-frequency sky. Multi-wavelength data shows that at least 94 per cent of the 10C sample are radio loud, we therefore suggest that these faint, flat-spectrum sources are the cores of radio galaxies. These source are split into HERGs and LERGs and their properties are compared in Whittam et al. (in prep). 

New data has enabled us to extend the 10C source count down to 0.1~mJy. We find no evidence for a change in the slope of the source counts, suggesting that the high-frequency sky is dominated by radio galaxies down to at least $S_{15.7} = 0.1~\rm mJy$. The source counts were compared to the S$^3$ and de Zotti et al.\ models, both of which under-predict the number of sources observed by a factor of two. This is probably because these simulations fail to model the dominance of the flat-spectrum cores of radio galaxies at 15.7~GHz. 

\acknowledgments

IHW and MJJ acknowledge financial support from the SKA South Africa. IHW acknowledges travelling support from the Ministry of Foreign Affairs and International Cooperation, Directorate General for the Country Promotion (Bilateral Grant Agreement ZA14GR02 - Mapping the Universe on the Pathway to SKA)

\end{document}